\documentclass[aps,twocolumn,prl,showpacs,superscriptaddress]{revtex4}

\usepackage{graphicx}
\usepackage{dcolumn}
\usepackage{bm}
\usepackage{epsfig}
\usepackage{pstcol,pst-grad}

\begin{document}

\title{Driving rate effects in avalanche-mediated, first-order phase
transitions}

\author{Francisco-Jos\'e   P\'erez-Reche}  
\affiliation{  Departament   d'Estructura i  Constituents  de la  Mat\`eria,
Universitat  de Barcelona  \\ Diagonal  647, Facultat  de F\'{\i}sica,
08028 Barcelona, Catalonia}
\author{Bosiljka  Tadi\'c}
\affiliation{Department   for   Theoretical   Physics,   Jozef   Stefan
Institute, P.O.Box 3000, SI-1001, Ljubljana, Slovenia.}
\author{Llu\'{\i}s Ma\~{n}osa}     
\author{Antoni Planes}     
\author{Eduard Vives}
\email{eduard@ecm.ub.es}   
\affiliation{  Departament   d'Estructura i  Constituents  de la  Mat\`eria,
Universitat  de Barcelona  \\ Diagonal  647, Facultat  de F\'{\i}sica,
08028 Barcelona, Catalonia}

\date{\today}

\begin{abstract}
We have  studied the  driving rate and  temperature dependence  of the
power-law  exponents that characterize  the avalanche  distribution in
first-order phase  transitions.  Measurements of  acoustic emission in
structural  transitions in  Cu-Zn-Al and  Cu-Al-Ni are  presented.  We
show how the observed behaviour  emerges within a general framework of
competing  time  scales of  avalanche  relaxation,  driving rate,  and
thermal  fluctuations. We  have  confirmed our  findings by  numerical
simulations of a prototype model.
\end{abstract}


\maketitle

The dynamics  of first-order phase transitions (FOPT)  in the presence
of  disorder is a  long-standing problem  which is  undergoing renewed
interest.   Besides  the phenomena  of  nucleation, metastability  and
hysteresis, additional features are presently being revisited: (i) the
extension  of  the transition  over  a  broad  region on  the  generic
temperature-field  phase diagram  (where  the field  is the  intensive
parameter conjugated to  the order parameter); and (ii)  the fact that
transitions  occur   through  a  sequence  of   avalanches  that  link
metastable states  when thermal fluctuations  are not dominant  at low
enough  temperatures.   These   order  parameter  discontinuities  are
associated with the sudden transformation of a fraction of the system.
This    behaviour    has    been    observed    in    many    magnetic
\cite{Babcock1990,Cote1991,Puppin2000,Durin2004},       superconducting
\cite{Wu1995,Altshuler2004}               and               structural
\cite{Vives1994a,PerezReche2001}    transitions    and    in    vapour
condensation  on porous  media \cite{Lilly1993}.   In many  cases this
phenomenon    has     been    described    as     ``athermal''    FOPT
\cite{PerezReche2001}.  In  practice, this means that for  the FOPT to
proceed it is necessary to  drive the system externally by varying the
temperature  $T$ or  the  generic field  $H$.   When $T$  and $H$  are
constant no avalanches  are detected even in the  case of long waiting
times.   This   reflects  the  fact  that   thermal  fluctuations  are
irrelevant  compared  with  the  high-energy barriers  separating  the
metastable states, and  that temperature acts as a  scalar field.  The
transformed fraction as a function of time only depends on the present
and past values of $T$ or $H$, but not explicitly on time.

In different systems, avalanches have  been detected with a variety of
experimental     techniques    including     induction    (Barkhausen)
\cite{Barkhausen1919,Durin2004},    magnetization    \cite{Hardy2004},
calorimetry  \cite{Carrillo1997}, resistivity  \cite{Wu1995}, acoustic
emission  (AE)  \cite{Vives1994a},  capacitance  \cite{Lilly1993}  and
optical       measurements      \cite{Babcock1990,Puppin2000,Kim2003}.
Interestingly,  in  many cases  the  statistical  distribution of  the
avalanche   properties  (area,   duration,  released   energy,  signal
amplitude, etc..)  exhibits  power-law behaviour over several decades.
This has suggested the existence of criticality and has stimulated the
search  for universality  in  the power-law  exponents. However,  many
different exponent  values have been reported and  comparison with the
theoretically          predicted          universality         classes
\cite{Sethna1993,Vives1995,Vives2001}         is        controversial.
Experimentally  accessible  quantities  such  as the  distribution  of
signal  amplitudes  are not  straightforwardly  comparable with  those
obtained in  numerical simulations such as the  avalanche sizes (which
are well  defined).  Moreover, the experimental  distributions in some
cases show  exponential corrections which make it  difficult to obtain
the power-law exponents.

Systematic measurements  with well-controlled external  parameters are
scarce.   In structural  transitions the  exponent associated  with AE
distributions has  been shown to depend on  repetitive cycling through
the  transition.   Stationary  values,   which  can  be  grouped  into
different  classes, are obtained  after a  relatively large  number of
cycles \cite{Carrillo1998}.  For  Barkhausen measurements in Fe-films,
an  increase  of the  exponent  has  been  found when  temperature  is
decreased  from room  temperature  to $10$  K \cite{Puppin2004}.   For
soft-magnetic    materials   two    classes    have   been    proposed
\cite{Durin2000}:  one   exhibiting  exponents  which   decrease  with
increasing driving  rate $r$  and the other  with exponents  which are
independent of $r$ \cite{Bertotti1994}.

In this Letter we present  systematic measurements of the influence of
cooling  rate $r=|\dot  T|$ on  the distribution  of amplitudes  of AE
signals in  structural transitions.  We analyze  two different systems
with different degree of athermal character. In one case, the exponent
has been found to {\it decrease} with $r$, while it {\it increases} in
the  other  case.   Results   are  well  explained  within  a  general
framework,  which accounts  for  the effect  of thermal  fluctuations.
Numerical simulations corroborate the proposed framework.

The   studied   samples   are  Cu$_{68.0}$Zn$_{16.0}$Al$_{16.0}$   and
Cu$_{68.4}$Al$_{27.8}$Ni$_{3.8}$    single    crystals   which    were
investigated  in a previous  work \cite{PerezReche2001}.   On cooling,
these alloys transform martensitically at $T_M=245$ K and $T_M=257$ K,
respectively.  Both  transitions extend in temperature  $\Delta T \sim
35$  K.   Previous results  \cite{PerezReche2001}  confirmed that  the
Cu-Zn-Al behaves athermally for  temperature driving rates larger than
a value which is less than $0.1$ K/min.  This was verified by checking
that the AE activity per temperature  interval as a function of $T$ is
independent  of $r$  (scaling).  In  contrast, Cu-Al-Ni  is  much less
``athermal''.   Although  it  clearly  shows avalanche  dynamics,  the
activity  does  not  scale  at  the  studied  driving  rates  and  few
isothermal  signals occur.   It was  estimated that  scaling  for this
sample will hold for $r$ above $50$ K/min.

\begin{figure}[ht]
\begin{center}
\epsfig{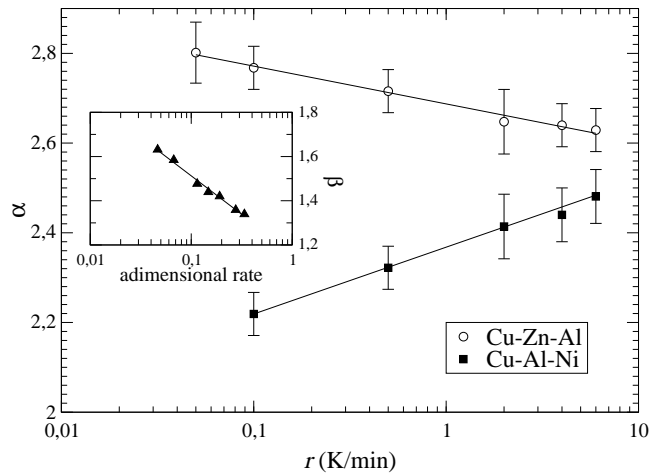}
\end{center}
\caption{\label{FIG1} Fitted exponents  corresponding to the amplitude
distribution of AE signals as a  function of the driving rate $r$. The
inset  shows  the exponents  corresponding  to  the Barkhausen  signal
integrated amplitudes from Ref. \onlinecite{Bertotti1994}}.
\end{figure}

The AE signals were detected  by a piezo-electric transducer using the
experimental  set-up explained  elsewhere  \cite{PerezReche2001}.  The
samples  were cycled  through the  transition more  than 100  times in
order  to  reach  the  stationary  path. Individual  AE  signals  were
detected during  cooling runs at  selected rates and  their amplitudes
$A$ were analyzed.  The experimentally  accessible range is $ 4 \times
10^{-5}$V$<A< 1.5 \times 10^{-3}$V.   For the Cu-Al-Ni sample averages
over  $\sim 10$ cycles  were also  performed in  order to  obtain good
statistics.  The typical  number of analyzed signals for  each rate is
$\sim 10^5$.  The temperatures  between which the cycles are performed
are always the  same: $T=220$ K and $T=340$ K  for the Cu-Zn-Al sample
and  $T=220$ K  and $T=320$  K for  Cu-Al-Ni. The  distribution  of AE
amplitudes  (as  well as  the  size  distribution  in the  simulations
presented below) is studied by  the maximum likelihood method which is
independent of  the way  histograms are plotted.   We have  fitted the
data  to  the  normalized  probability  law:  $p(A)  \sim  A^{-\alpha}
e^{-\lambda  A}$.    Two  different  fits  are   performed:  first,  a
one-parameter  fit of  the exponent  $\alpha$ by  imposing $\lambda=0$
and,  second, a  two-parameter  fit of  $\alpha$  and $\lambda$.   The
values of $\alpha$ reported here correspond to the first fitted value,
but error  bars include the second  fitted value.  For  the second fit
$|\lambda|<  1000$  V$^{-1}$  in   all  cases,  which  indicates  that
differences  between the two  fitted functions  start to  be important
outside  of  the experimental  window.   Figure  \ref{FIG1} shows  the
dependence of $\alpha$ on the  rate $r$.  The striking feature is that
for  the most  athermal case  (Cu-Zn-Al) $\alpha$  decreases  with $r$
while  for   the  less  athermal  sample   (Cu-Al-Ni),  it  increases.
Moreover,  in both  cases, the  dependence on  $r$ is  logarithmic, as
shown by the  fitted lines.  A decrease of  the exponent of Barkhausen
avalanche sizes (area below the signals) in polycrystalline Si-Fe with
increasing  rate is  reported  in Ref.~\onlinecite{Bertotti1994}.   We
have  plotted  these  data on  a  log-linear  scale  in the  inset  of
Fig.~\ref{FIG1}, and a good logarithmic dependence is also observed.

Understanding of the  results comes from the analysis  of the relevant
time scales involved in the problem.  These are: (i) The time scale of
the  avalanche relaxation  $\tau_{av}$.  Although  avalanche durations
are known to  be also power-law distributed over  several decades, the
detected values  range between $\tau_{av}  \sim 10^{-6} -  10^{-3}$ s.
(ii)  The  characteristic  time   associated  with  the  driving  rate
$\tau_{dr}$.  This  can be conveniently  defined as a function  of the
extension of the FOPT as $\tau_{dr}=\Delta T/r$.  In our case from the
experimentally  accessible  rates  $\tau_{dr}  \sim 10^{2}  -  10^{4}$
s. (iii) The third time scale is associated with the activation of the
FOPT by  thermal fluctuations $\tau_{th}$.  This  scale decreases with
temperature and depends on the  free energy barriers. For our samples,
given the  above-mentioned limiting rates for  the athermal behaviour,
we have estimated that $\tau_{th}> 10^5$ s for Cu-Zn-Al and $\tau_{th}
\lesssim  10^2$ s for  Cu-Al-Ni.  Moreover,  given that  $\Delta T/T_M
\sim 0.1$  is small for the  studied samples, we expect  that, in both
cases, $\tau_{th}$ does not depend on $r$.

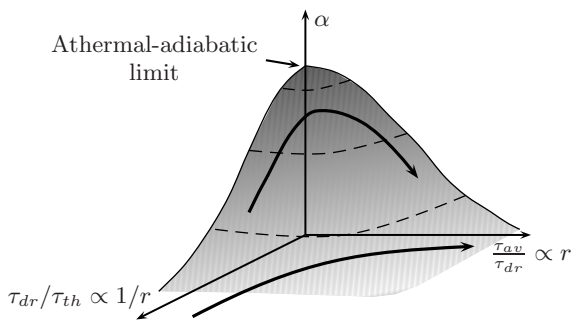
\begin{figure}[ht]
\begin{center}
\begin{pspicture}(-2.625,-0.9)(3,2.9)
\psset{xunit=0.75,yunit=0.75}

\pspolygon[fillstyle=gradient,%
 linestyle=none,%
 gradbegin=darkgray,%
 gradend=white,%
 showpoints=false,%
 gradangle=10]%
(0,3)(0.5,2.9)(1,2.6)(1.5,2.1)(2,1.6)(2.5,1)(2.84,0.7)(3.3,0.4)(3.6,0.2)%
(3.8,0.1)(2,-0.9)(1.6,-1.05)(1.2,-1.1)(0.5,-1.075)(0,-1.06)(-1,-1.025)
(-2.6,-1)(-2.2,-0.6)(-2,-0.35)(-1.6,0.25)(-1.35,0.7)(-1,1.5)(-0.5,2.5)(-0.25,2.8)

\psline[linewidth=0.75pt]{->}(0,0)(-3,-1.5)
\rput(-4.0,-1.1){$\tau_{dr}/\tau_{th} \propto 1/r$}
\psline[linewidth=0.75pt]{->}(0,0)(4,0)
\rput(4,-0.4){$\frac{\tau_{av}}{\tau_{dr}} \propto r$}
\psline[linewidth=0.75pt]{->}(0,0)(0,4)
\rput(0.3,3.8){$\alpha$}

\pscurve[showpoints=false,linewidth=0.5pt](0,3)(0.5,2.9)(1,2.6)(1.5,2.1)(2,1.6)(2.5,1)(2.84,0.7)(3.3,0.4)(3.6,0.2)(3.8,0.1)

\pscurve[showpoints=false,linewidth=0.5pt](0,3)(-0.25,2.8)(-0.5,2.5)(-1,1.5)(-1.35,0.7)(-1.6,0.25)(-2,-0.35)(-2.2,-0.6)(-2.6,-1)

\psbezier[showpoints=false,linewidth=1.2pt]{->}(-2,-1.45)(0,-0.4)(1,-0.3)(3,-0.2)
\pscurve[showpoints=false,linestyle=dashed,linewidth=0.5pt](-0.4,2.6)(0.1,2.57)(0.8,2.75)
\pscurve[showpoints=false,linestyle=dashed,linewidth=0.5pt](-1,1.5)(0.4,1.45)(1.8,1.8)
\pscurve[showpoints=false,linestyle=dashed,linewidth=0.5pt](-1.6,0.1)(0.7,0)(2.84,0.7)

\pscurve[showpoints=false,linewidth=1.2pt]{->}(-1,0.4)(-0.75,0.9)(0,2.1)(0.2,2.2)(1,2)(2,1)

\psline{->}(-0.7,3.1)(-0.1,3)
\rput[r](-0.75,3.1){\begin{tabular}{c} \small{Athermal-adiabatic} \\ \small{limit}\end{tabular}}

\end{pspicture}
\end{center}
\caption{\label{FIG2} Schematic  diagram showing the  behaviour of the
exponent as a function of  the relevant time scales.  The hyperbola on
the  base  plane shows  the  behaviour  at  constant temperature  when
increasing the driving rate.}
\end{figure}

The dynamics of  the FOPT will therefore be  determined by competition
between   these  three   time  scales.    In  the   "adiabatic"  limit
$\tau_{dr}/\tau_{th}\rightarrow     0$     and    $\tau_{av}/\tau_{dr}
\rightarrow  0$,  i.e.,  when  the  time  scales  are  well  separated
$\tau_{av}  \ll  \tau_{dr} \ll  \tau_{th}$,  the  avalanches are  well
defined and  both temperature  and finite driving  rate do  not affect
avalanche  scaling.   When  $\tau_{dr}$  approaches  $\tau_{av}$,  the
system  will still  display  avalanche behaviour  but,  will start  to
overlap.   Typically, small  avalanches merge  to form  larger signals
\cite{White2003}.  This  leads to a decrease in  the exponent $\alpha$
when $\tau_{av}/\tau_{dr} \propto r$ increases.  On the other hand, by
increasing temperature and or  decreasing driving rate, competition of
the  thermal fluctuations  and  driving may  produce diverse  effects.
Strictly   speaking,   for   any    nonzero   value   of   the   ratio
$\tau_{dr}/\tau_{th} \propto  1/r$ the activity  may not stop  even at
$r=0$  until thermal  equilibrium is  reached, making  the  concept of
avalanches difficult  to define.  Nevertheless,  from the experimental
point of view, there is  always a threshold below which signals cannot
be detected.   Thus, distinct  signals can still  be measured  for low
enough  values of  $\tau_{dr}/\tau_{th}$.  We  expect  that increasing
thermal activation  (decreasing $\tau_{th}$) will help  the advance of
the FOPT  by promoting the  transformation of domains  which otherwise
would not transform under {\it athermal} circumstances, thus resulting
again in the merging of  small avalanches.  On the other hand, thermal
fluctuations may  initiate small avalanches when the  system is slowly
driven ($\tau_{dr}$  is large) which  will not initiate at  such early
fields when  the system is driven  fast. The potential  effects of the
two time scale  ratios $\tau_{av}/\tau_{dr}$ and $\tau_{dr}/\tau_{th}$
on the  scaling exponents are shown  schematically in Fig.~\ref{FIG2}.
Generally,  for a  given system  at finite  temperature,  driving rate
dependence is determined along the projected curve on the basal plane.
Thus,  depending on  the  experimentally accessible  range of  driving
rates we  can find an increase  of the exponent for  the less athermal
samples, a region of constant exponents, or a decrease of the exponent
for  the more  athermal samples.   This enlarged  space  of parameters
offers a suitable  scenario to understand the present  results for the
acoustic emission  measurements, as well  as the behavior  observed by
large  temperature changes  recently reported  \cite{Puppin2004}.  The
studied  case of  Barkhausen  avalanches \cite{Bertotti1994,Durin2000}
represents the limit where thermal fluctuations can be neglected.

In  order  to  substantiate  the  above arguments  we  have  performed
numerical  simulations.   The   prototype  model  for  hysteresis  and
avalanches in athermal FOPT is  the 3D-Random Field Ising Model (RFIM)
at $T=0$ with metastable dynamics \cite{Sethna1993}.  It consists of a
ferromagnetic Ising model with  local random fields which are Gaussian
distributed with  standard deviation  $\sigma$.  The system  is driven
from the saturated state by  decreasing the external field $H$.  Spins
flip  according to  a local  relaxation rule  which is  reponsible for
hysteresis.  By  construction, the model  is in the  perfect adiabatic
limit  $\tau_{av}\ll  \tau_{dr}   \ll  \tau_{th}=\infty$.   The  first
inequality is guaranteed because as  soon as one avalanche starts, $H$
is kept constant until the  system reaches a new metastable state.  We
will refer to such  dynamics as {\it athermal-adiabatic} dynamics. The
model  reproduces  the existence  of  avalanches  whose  size $s$  and
duration $\Delta t$ can be  statistically analyzed.  In a large region
close    to    the    critical    point   $\sigma_c    \simeq    2.21$
\cite{PerezReche2003},  the avalanche  sizes corresponding  to  a full
half-loop  distribute,  to  a   good  approximation,  according  to  a
power-law.   Sensitivity  of  the  exponent  to system  size  $L$  and
$\sigma$     has      recently     been     studied      in     detail
\cite{PerezReche2003,PerezReche2004b}.    Moreover,  given   that  the
distributions  are  only approximate  power-laws,  the exponents  also
depend on the fitting range.

\begin{figure}[ht]
\begin{center}
\epsfig{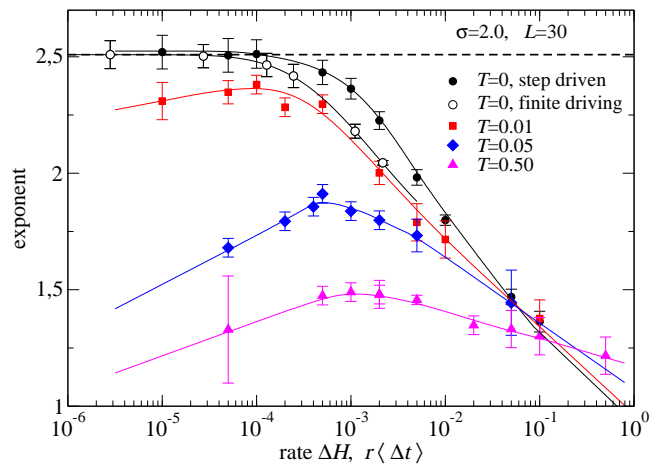}
\end{center}
\caption{\label{FIG3}  Exponents corresponding  to the  avalanche size
distribution as a  function of the driving rate  obtained by different
numerical algorithms  at different  temperatures $T$, as  explained in
the text.}
\end{figure}

Extension of the model in order to incorporate the finite driving rate
$r=|\dot H|$ and thermal  fluctuations effects is not straightforward.
The appropriate enlargement of the  parameter space can be achieved in
several ways.  Here  we adapt the approach in  which the separation of
signals  is  preserved  to  a   large  degree.   A  first  attempt  to
incorporate  the   finite  driving  rate,  which  we   refer  to  {\it
athermal-step-driven}     dynamics,    was    proposed     in    Refs.
\onlinecite{Tadic1996,Tadic2002}.  (Here we  extend this work to lower
driving rates). It consists of  increasing the external field $H$ by a
certain step  $\Delta H$ and keeping  it fixed until  a new metastable
situation is  reached. The obtained signals are  superpositions of the
avalanches of the  athermal-adiabatic model.  The statistical analysis
shows  that  the  exponent  decreases  with  the  increasing  ``rate''
(quantified by  $\Delta H$).  Results  for a system  with $\sigma=2.0$
are shown in Fig.~\ref{FIG3} as filled circles
\footnote{The  actual  values  of   the  exponent  obtained  with  the
different  numerical  algorithms  discussed  in this  work  depend  on
$\sigma$, $L$ and the fitting range of $s$.  We have verified that for
$2.0<\sigma<2.4$ and $L>20$,  the qualitative behaviour for $2<s<1000$
is  that  shown  in  Fig.~\ref{FIG3}}.   Data are  compared  with  the
exponent corresponding  to athermal-adiabatic dynamics  which is shown
by  a dashed  horizontal line.   A  logarithmic decrease  with $r$  is
obtained in a broad region,  in qualitative agreement with the results
obtained for Cu-Zn-Al \footnote{The actual values of the exponents are
not comparable since the simulations correspond to avalanche sizes and
the experimental data correspond to amplitudes of AE signals.}.

Strictly speaking,  for the athermal-step-driven dynamics,  $r$ is not
constant  along the hysteresis  path.  This  could be  important since
long avalanches  are known to concentrate  on the central  part of the
hysteresis loop and correlations may exist.  In order to check whether
or not  this affects the results,  we propose other  dynamics which we
will  call  {\it  athermal-finite-driving-rate} dynamics.   Using  the
original  athermal-adiabatic  dynamics, one  records  the sequence  of
avalanches in  half a loop (including  the field value  for which they
occur, their size, and their duration).  For a given finite rate $\dot
H$, it is possible to  analyze the sequence of recorded avalanches and
determine  which of  them  will  overlap.  Thus,  one  can obtain  the
modified  sequence of  signals and  perform the  statistical analysis.
Results obtained  by this method  are indicated in  Fig.~\ref{FIG3} as
open circles.  In order to compare it with the previous data, $\dot H$
has been multiplied by the average duration $\langle \Delta t \rangle$
of the pulses in a half loop.  Agreement with the previous dynamics is
very  good,  thus  supporting  the  use  of  step-driven  dynamics  in
simulations.

The  numerical results obtained  are in  agreement with  the behaviour
proposed     in    Fig.~\ref{FIG2}     and    correspond     to    the
$\tau_{dr}/\tau_{th}=0$  plane.  To  simulate the  effect of  a finite
$\tau_{th}$ we must modify the dynamics proposed above introducing the
effect of  thermal fluctuations,  but still keeping  separate signals.
This   is    achieved   by   slightly    modifying   the   step-driven
algorithm. After each  $\Delta H$, not only unstable  spins relax, but
also  an extra  small fraction  of  locally stable  spins is  reversed
towards the new  phase due to thermal activation.   This is done ``\`a
la Metropolis'', simultaneously and  independently for each spin, with
a  probability $p=\min(1,e^{-\Delta  E/T})$, where  $\Delta E$  is the
energy change associated with the reversal of each spin.
During the  subsequent evolution of  the avalanche, the field  is kept
constant  and temperature  is set  to  $0$.  The  exponents for  three
different temperatures  are shown in Fig.~\ref{FIG3}.  As  can be seen
non-monotonous behaviour is  obtained, as qualitatively illustrated in
Fig.~\ref{FIG2}.

Several models have been  previously proposed for the understanding of
the   influence   of  driving   rate   on   the  power-law   exponents
characterizing                   signal                  distributions
\cite{Alessandro1990,Durin2004,Queiroz2001,White2003}.   All  of  them
correspond to athermal situations and  account for the decrease of the
exponent with increasing $r$, as a consequence of the overlap of small
avalanches.  In some cases a linear decrease of the exponents with $r$
is predicted.  Our experiments and simulations fit better with a $\log
r$ dependence.  Moreover,  we have also shown that  Barkhausen data in
Ref.  \cite{Bertotti1994}  also conform to  this logarithmic tendency.
We expect  that the origin of  this functional dependence  lies in the
correlation between avalanches evolving  in the restricted geometry of
the  transforming sample  rather than  in the  dynamics  itself.  Such
correlations have not been taken into account in previous models which
correspond    to    single    advancing    interfaces    in    magnets
\cite{Alessandro1990,Durin2004,Queiroz2001}     or     systems    with
uncorrelated avalanches \cite{White2003}.

In conclusion, we have shown that the avalanches at FOPT may exhibit a
variety  of the  driving  rate dependences  related  to the  interplay
between the three relevant time scales.  We have, therefore, clarified
under  which  experimental  circumstances  one  can  expect  exponents
similar to those predicted in the athermal adiabatic limit.

We  thank A.Saxena  for  fruitful comments.   This  work has  received
financial support  from CICyT (Spain), project  MAT2001-3251 and CIRIT
(Catalonia),  project 2001SGR00066.   B.T.  acknowledges  support from
DURSI  (Catalonia)  and  project  No.   P1-0044  of  the  Ministry  of
Education,  Science   and  Sports  (Slovenia).F.J.P.-R.   acknowledges
support from DGICyT (Spain).

\end{document}